\documentclass{article}

 \usepackage{graphicx}

\usepackage{amssymb}

\begin{document}





\title{Neutron diffraction constraint on spin-dependent short range interaction}


\author{V.V. Voronin, V.V. Fedorov, I.A. Kuznetsov}

\maketitle
\begin{center}
Petersburg Nuclear Physics Institute,\\ 188300, Gatchina,
St.Petersburg, Russia\\
\end{center}

\begin{abstract}
The direct constraint on the parameters of short range pseudomagnetic interaction of free neutron with matter is obtained from the recent test experiment on a search for neutron EDM by crystal-diffraction method \cite{dedm_test}. 
It is shown that this constraint on a product of scalar to pseudo-scalar coupling constants $g_s g_p$ is better than that of any other method for the range $\lambda < 10^{-5}$cm. 
\end{abstract}

\section{Introduction}
Over last years a possibility to look for new hypothetic particles which results in a new short range Yukawa-type potential of fermion-fermion interaction is actively discussed. The spin-dependent short-range interactions may be induced by light, pseudoscalar bosons such
as the axion invented to solve the strong CP problem \cite{Moody}.  This interaction is usually parameterised as \cite{Moody}
\begin{equation}
	V_{SP}({\bf r})=\frac{\hbar^2 g_S g_P}{8 \pi m}\left(\frac{\bf r}{r}\cdot \mbox{\boldmath $\sigma$}\right)\left(\frac{1}{r\lambda}+\frac{1}{r^2}\right)e^{-r/\lambda}
\label{eq:V_SP}
\end{equation}
where $g_S$ and $g_P$ are nondimensional parameters of the scalar and pseudo-scalar coupling constants between the neutron and exchanged boson. $\lambda=\hbar /m_Ac$ is typical parameter of the range of forces (Compton wavelength of axion). 
There are proposals to search this new type of interaction using gravitationally bound quantum states of a free neutron \cite{Nesv1} and using a spin precession of the trapped ultracold neutrons in vicinity of bulk matter \cite{Oliver1}. Both of these methods have a suitable sensitivity for the range of $\lambda>10^{-3}$cm, but their sensitivity is extremely decreased for the range $\lambda<10^{-4}$cm. 

Here we consider a possibility to use a neutron diffraction in the perfect non-centrosymmetric crystal to search a new type of short range interaction for the $10^{-10}<\lambda<10^{-5}$cm. 

\section{Diffraction in a non-centrosymmetric crystal}

Neutron diffraction in a non-centrosymmetric crystal was widely discussed within the framework of the project to search for the neutron electric dipole moment (nEDM) by the diffraction method \cite{dedm,Dedm2}.

Any crystal potential (nuclear, electric, new short range potential, ..) can be presented as sum of the potentials of  different atoms placed into the crystal cell. For the periodic crystal structure it is convenient to present such potential as Fourier series over the reciprocal lattice vectors $\mbox{\boldmath $g$}$  

\begin{equation}
 V(\mbox{\boldmath $r$}) = \sum_a V_a(\mbox{\boldmath $r-r_a$})=
  \sum_g V_g e^{i\mbox{\boldmath $gr$}}=
   V_0 + \sum_g 2v_g\cos(\mbox{\boldmath $gr$}+\phi_g),
\label{eq:2v(r)}
\end{equation}
where $V_a(\mbox{\boldmath $r-r_a$})$ is the potential of single atom, {\boldmath $r_a$} is the atom position, $V_g = v_g \exp(i\phi_g)$, $g=2\pi/d$, $d$ is the interplanar distance. 
Here we take into account $V_g=V_{-g}^*$, because we consider the real value potentials.

g-harmonics of potentials can be found from the equation
\begin{equation}
 V_g=\int\limits_{v=1} d^3r~e^{-i\mbox{\boldmath $gr$}}V(\mbox{\boldmath $r$}),
\label{eq:(3)}
\end{equation}

In the case of nuclear potential 
\begin{equation}
 V_g= -\frac{2\pi\hbar^2}{m V_c}F_g,
\label{eq:(31)}
\end{equation}
here $m$ is the neutron mass, $V_c$ is volume of crystal unit cell, $F_g$ is the structure amplitude
\begin{equation}
  F_g=
\sum_i e^{-W_{ig}}f_i(\mbox{\boldmath $g$})e^{-i\mbox{\boldmath $gr$}_i}.
\label{eq:(4)}
\end{equation}
Here we sum over the atoms of unit cell,
$f(\mbox{\boldmath $g$})$ is the scattering amplitude of $i$ atom, $W_{ig}$
is the Debye-Waller temperature factor.

For the case of a non-centrosymmetric crystal the different potentials can be shifted to each others, by the other words, the phases $\phi_g$ of different crystal potentials can be not equal.  For the case of electric potential this shift results in a large electric field affected the neutron in non-centrosymmetric crystal \cite{dfield,Dedm1}. We take the phase of the nuclear potential equal to zero so 
the value of the electric field affected the neutron will be
\begin{equation}
 \mbox{\boldmath $E$}(\mbox{\boldmath $r$}) =
 -\mbox{grad}~V^E_g(\mbox{\boldmath $r$})=
  2v^E_g\mbox{\boldmath $g$}\sin(\mbox{\boldmath $gr$}+\phi^E_g).
\label{eq:9}
\end{equation}
where $v^E_g$ and $\phi^E_g$ are amplitude and phase of the g-harmonics of crystal electric potential accordingly. 
Let's consider the monopole-dipole interaction (\ref{eq:V_SP}).
Direct calculation of g-harmonic of $V_{SP}({\bf r})$ from (\ref{eq:(3)}) gives 

\begin{equation}
\hat{V}^{SP}_g=-i F^{SP}_g  e^{i\Phi^{SP}_g}{{ \frac{\hbar ^2 g_s g_p }{2m V_c}}}{{ \frac{g \lambda^2 }{1 + g^2 \lambda^2 }}}(\mbox{\boldmath $\sigma$} {\bf n}_g)
\end{equation}
where ${\bf n}_g\equiv \mbox{\boldmath $g$}/g$, $F^{SP}_g $ and $\Phi^{SP}_g$ are the amplitude and phase of structure factor $f^{SP}_g $ of the crystallographic plane.
$f^{SP}_g $ is determined by the following sum 
\begin{equation}
f^{SP}_g = \sum_{i}A_i\cdot e^{i {\mbox{\boldmath $g r$}_i}}
\end{equation}
here $A_i$ and ${\bf r}_i$ is the mass and position of a corresponding atom in elementary cell.

Neutron wave function is determined by the nuclear interaction, and for the case of close to Bragg direction passage through the crystal can be written as \cite{Dedm2,Dedm3}

\begin{equation}
    \psi({\bf r})= e^{i\mbox{\boldmath $kr$}}+\frac{V_g^N}{E_k-E_{k_g}}
        e^{i\mbox{\boldmath $k_g r$}}\equiv e^{i\mbox{\boldmath $kr$}}
     \left[1 - \frac{U_g^N}{2\Delta_g}e^{i\mbox{\boldmath $gr$}}\right],
\label{eq:23}
\end{equation}
where we take into account that the phase of nuclear harmonics  $V_g^N$ equal to zero,  {\boldmath $k_g$}={\boldmath $k+g$},
$E_{k}= \hbar^2 k^2/2m$, $E_{k_g} = \hbar^2 k_g^2/2m $, $V_g^N = \hbar^2 U_g^N
/2m$, $\Delta_g = (k^2_g - k^2)/2$ is the parameter of deviation from the exact Bragg condition.

Therefore, $\hat{V}_{SP}$ potential affecting the neutron in the crystal will be 
\[
\hat{V}_{SP}=\langle\psi({\bf r})|V_{SP}({\bf r})|\psi({\bf r})\rangle = \frac{U_g^N}{\Delta_g} |\hat{V}^{SP}_g|\sin\Phi^{SP}_g = 
\]

\begin{equation}
=\frac{U_g^N}{\Delta_g} F^{SP}_g {{ \frac{\hbar ^2 g_s g_p }{2m V_c}}}{{ \frac{g \lambda^2 }{1 + g^2 \lambda^2 }}}(\mbox{\boldmath $\sigma$} {\bf n}_g)\sin\Phi^{SP}_g \equiv V_{SP}(\mbox{\boldmath $\sigma$} {\bf n}_g).
\label{eq:v_sp}
\end{equation}

We should note that for centrosymmetric crystal $\Phi^{SP}_g\equiv 0$ and in this case the mean potential affecting the neutron will be zero. One can see also from (\ref{eq:v_sp}) that this "`pseudomagnetic"' potential is  proportional to the  parameter $\Delta_B\equiv U_g /\Delta_g$ determined by the deviation from the Bragg condition. That allows  to control the value and sign of the potential selecting the neutrons with slightly different energies from the Bragg one. 

Interaction with such a potential will lead to the neutron spin rotation around the reciprocal lattice vector $\mbox{\boldmath $ g$}$ by the angle

\begin{equation}
\varphi_{SP}=\frac{2V_{SP}}{\hbar} \tau
\end{equation}
where $\tau$ is the time of neutron travel through the crystal.

\section{The method sensitivity}

For example let's consider (110) plane of non-centrosymmetric quartz crystal and $\Delta_B=0.5$.
For (110) plane $g=2.56\cdot 10^8$cm$^{-1}$, $F^{SP}_g=51$, $\sin(\Phi^{SP}_g)=0.41$, $V_c=113 $\AA$^3$.
The angle of spin rotation due to considered potential will be
\begin{equation}
\varphi_{SP}= 0.36\cdot 10^{24} [cm^{-3}] \cdot \frac{g_S g_P }{g^2+1/\lambda^2} L
\end{equation}
where $L$ is the crystal length.
For the cold neutron beam at high flux reactor the measurement accuracy $\sigma(\varphi_{SP})\sim 2 \cdot 10^{-6}$ can be reached for 100 day of the statistic accumulation \cite{dedm_test}. That allows to give the constraint for monopole-dipole interaction 
\begin{equation}
g_S g_P <  10^{-31} [cm^{2}]\cdot ({g^2+1/\lambda^2})
\end{equation}
for the L=50 cm.

\begin{figure}[htbp]
	\centering
		\includegraphics[width=0.8\textwidth]{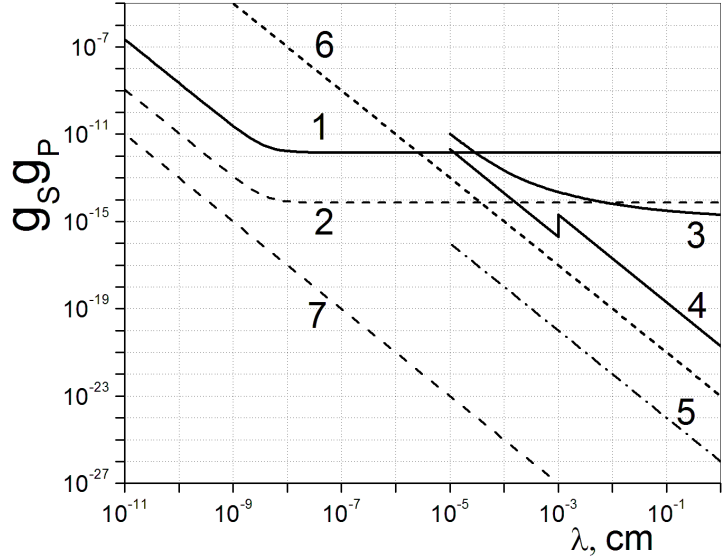}
	\caption{Constraints on a value of coupling constants product $g_s g_p$. Curve (1) is the constraint from the crystal-diffraction nEDM experiment \cite{dedm_test} (this work) and (2) is possible improvement of this method, (3) is gravitational level experiment \cite{Nesv1}, (4) is the UCN depolarization \cite{Serebrov}, (5) is proposal \cite{Oliver1}, (6) and (7) are the predictions of axion model with $\theta\sim 1$ and $\theta\sim 10^{-10}$ correspondingly \cite{Serebrov,Raffelt}}
	\label{fig:sensitivity}
\end{figure}

Recently the test experiment for the search for neutron EDM by crystal-diffraction method was carried out \cite{dedm_test}. This result already allows to give the direct constraint on a value of $g_s g_p$ better than any other method for the $\lambda < 10^{-5}$cm, see  Fig.\ref{fig:sensitivity}, curve (1).

The comparison of different constraints on $g_S g_P$ is shown in Fig.~\ref{fig:sensitivity}. 

One should note that both neutron EDM interaction with crystal electric field and the spin-dependent short-range interaction lead to a neutron spin rotation about reciprocal lattice vector $g$. Therefore, the considered shot-range interaction can give a false effect for the neutron EDM experiment and vice versa. However, these two interactions will be different for different crystallographic planes, so in the case of nonzero effect they can be separated using different planes for measurement.  

\section{Conclusion}
Direct constraint on amplitude of T-odd monopole-dipole interaction of neutron with the matter was obtained. It is shown that the product of scalar to pseudo-scalar coupling constant $g_s g_p<10^{-12}$ for the range $10^{-8}<\lambda< 10^{-5}$cm. This value can be improved on about $10^3$ times for the full scale setup for the neutron EDM search by crystal-diffraction method, which is under construction now.  

This work is supported by grant RFBR-09-02-00446.

\end{document}